\documentstyle[art12,12pt,epsf]{article}
\begin{document}

\renewcommand{\thefootnote}{\fnsymbol{footnote}}

\begin{titlepage}
  \begin{flushright}
SLAC-PUB-8211\\
 July, 1999\\
  \end{flushright}

\vskip 1.5cm

\begin{center}

\Large{\bf  Combined SLD Measurement of $A_b$ at the $Z^0$ Resonance
using Various Techniques.\footnote{Work supported in part by the
Department of Energy contract  DE--AC03--76SF00515.}}
\vskip 0.3cm

The SLD Collaboration$^{**}$


\end{center}

\vglue 2.0cm
\begin{abstract}
\noindent

We present a new preliminary combination of measurements of the 
parity-violation parameter $A_b$
made by the SLD collaboration using various experimental techniques.
The techniques differ in detail, but in general  
a sample of $b\bar{b}$ events is selected or enhanced by
using the topologically reconstructed mass of the separated
vertices formed by decaying $B$ hadrons.
The direction of the $b(\bar{b})$ quark is signed by one
of four final state tags: jet charge, vertex charge, leptons, 
or identified $K^{\pm}$ from the $b$ vertex. We account for
statistical and systematic correlations between the four analyses
to arrive at our combined result:
$A_b=0.905 \pm 0.017 (\mbox{stat}) \pm 0.020 (\mbox{syst})$.

\end{abstract}

\vglue 3.cm
\centerline{\it Paper Contributed to the International Europhysics Conference on High Energy}
\centerline{\it Physics (EPS-HEP 99), July 15-21, 1999, Tampere, Finland, Ref 6-473;}
\centerline{\it and the XIX International Symposium on Lepton and Photon Interactions }
\centerline{\it  at High Energies, August 9-14, 1999, Stanford, CA, USA}

\end{titlepage}

This note provides a brief overview of four separate analyses performed
by the SLD Collaboration to measure  the parity-violation parameter
$A_b$ in polarized $Z^0$ decays, and a description of how the analyses
are combined to form a overall SLD result. The reader is referred to the
detailed notes available for each analysis for specific information on
how each analysis is performed.

The most statistically powerful analysis selects $b\bar{b}$
events using an inclusive topological vertexing technique\cite{ZVTOP}
and forms the momentum-weighted jet charge of all selected events 
to identify the quark direction\cite{JETQ98}. This analysis was most 
recently updated
at Moriond '99 to include the full 1993-8 SLD dataset\cite{MORIOND99}. 
The updated systematic errors are reproduced in Table~\ref{systab_jetq}.
The combined jet-charge result is:
\begin{equation}
  A_b=0.882\pm 0.020(\mbox{stat})\pm 0.029(\mbox{syst})\quad (\mbox{\bf jet-charge}) .
\end{equation}

The next analysis uses identified high-momentum muons and electrons 
to tag heavy flavor ($b,c$) events and then employs a number of kinematic and 
vertexing variables to try to distinguish
leptons arising from $b$-hadron decays from those arising from $c$-hadron
decays. The lepton sign is used to sign the quark direction and $A_b$ and
$A_c$ are measured simulataneously\cite{LTAG98}. This analysis was most 
recently updated at Moriond '99 to include the full 1993-8 SLD 
dataset\cite{MORIOND99}.  
The updated systematic errors are reproduced in Tables~\ref{systab_mu}
($\mu^{\pm}$ tag) and ~\ref{systab_el} ($e^{\pm}$ tag).
The combined lepton-tag result is:
\begin{equation}
  A_b=0.924\pm 0.032(\mbox{stat})\pm 0.026(\mbox{syst})\quad (\mbox{\bf leptons}) .
\end{equation}

Another analysis uses identified $K^{\pm}$ associated with separated
topological vertices to sign the quark direction, exploiting the 
dominant $b\rightarrow c\rightarrow s$ decay chain.
In the original version of this analysis\cite{KTAG98},
the error in the result was dominated by the experimental
uncertainty in the relative rates of $B\rightarrow K^+ X$ vs.
 $B\rightarrow K^- X$ decay. This analysis has been updated
at this conference\cite{KTAGNEW} to include data from the 1997-8
data run and now employs a self-calibration technique which
removes the reliance on relative production rates of
$K^{\pm}$ in $B$ decays. 
The combined $K^{\pm}$-tag result is:
\begin{equation}
  A_b=0.960\pm 0.040(\mbox{stat})\pm 0.056(\mbox{syst})\quad (\mbox{\bf kaons}) .
\end{equation}

The last analysis uses the charge of the separated topological
vertices themselves to assign the quark direction.  
The vertex charge is weighted in the analysis based on the
mass of the reconstructed vertex, which gives an indication of
the fraction of the $B$ decay tracks which have been correctly
assigned to the vertex.
This analysis, which has been first presented
at this conference\cite{VTXCHG}, includes data from the 1996-8
data run and also employs a self-calibration technique 
to determine the correct-sign probability directly from the data. 
The vertex-charge result is:
\begin{equation}
   A_b=0.897\pm 0.027(\mbox{stat})^{+0.036}_{-0.034}(\mbox{syst})\quad (\mbox{\bf vertex-charge}) .
\end{equation}

We have combined these four results as follows. The statistical overlap
between the analyses was determined by explicitly tabulating events
used by the four analyses for a subset of the total data which
is common to all four and was marked by stable detector performance.
Each event in this dataset used by a given analysis was assigned 
a weight by that analysis based on its estimated $b$-hadron purity,
correct-signing probability, and reconstructed polar angle. 
The statistical correlations between analyses for this dataset was 
then determined from the overlapping event fractions, the fractions
of events where different tags assigned the same (opposite) quark
directions, and the individual event weights. This statistical
correlation was then diluted to account for the fact that not all
analyses use the same dataset.  

The statistical correlations extracted range from $\sim 10-30\%$
depending on the pair of analyses considered. The largest correlation
(28\%) was observed between the jet-charge and vertex-charge analyses,
as expected; due to its statistical power the jet-charge analysis
has significant overlap with all three other analyses. The smallest
correlation (8\%) was between the lepton tag and vertex charge 
analyses. 

Correlations between analyses due to common systematic error sources
have been treated in the standardized fashion developed by the LEP
Electroweak Working Group \cite{LEPEWWG}. Since three of the four 
analyses (all but the lepton tag) use self-calibration techniques 
based on the data, most of the quoted systematic errors are in fact
dominated by data statistics and thus (mostly) uncorrelated.
For the purposes of this combination, we assume $A_c$ is fixed at its
Standard Model value. 

The analyses are then combined in a weighted average
using the individual analysis errors and the statistical correlation
matrix. Each analysis receives a weight in the overall combination
based on its statistical and uncorrelated systematic error. 
Statistical and uncorrelated systematic errors are combined in
quadrature and correlated systematic errors are combined linearly.
The final analysis weights are 38\% (jet-charge), 30\% (leptons),
22\% (vertex-charge), and 10\% (kaons).
The combined SLD preliminary result obtained with this procedure is:
\begin{equation}
   A_b=0.905\pm 0.017(\mbox{stat})\pm 0.020(\mbox{syst})\quad (\mbox{\bf combined}) .
\end{equation}
This result differs slightly from the LEP Electroweak Working Group
fit \cite{MNICH, MOENIG} of the same data due to correlations between 
the $A_b$ and $A_c$ results which enter here primarily through the 
lepton-tag analysis. We explicitly ignore such correlations in
our average whereas the LEP global fits include them.   

Our average result for $A_b$ agrees well with the Standard Model expectation
of 0.935, and also with that derived from the current combination of LEP 
results ($0.892 \pm 0.024$) used in the global electroweak fit\cite{MNICH}. 
The combined LEP and SLD results, however, imply that $A_b$ deviates from 
the Standard Model at the $\sim 2.5\sigma$ level; this intriguing situation
has persisted since 1996 despite significant improvements in statistical
and systematic errors. One recent analysis \cite{FIELD} of the world's 
$A_b$ data shows no evidence of systematic bias or underestimated errors.
Thus the experimental question of possible anomalies in the $Zb\overline{b}$
coupling remains unresolved.

\vfill\eject

\begin{center}
{\Large \bf The SLD Collaboration}
\def\iADEL{$^{(1)}$}
\def\iAOMORI{$^{(2)}$}
\def\iBOLO{$^{(3)}$}
\def\iBRI{$^{(4)}$}
\def\iBRUN{$^{(5)}$}
\def\iBU{$^{(6)}$}
\def\iCINC{$^{(7)}$}
\def\iCOLO{$^{(8)}$}
\def\iCOLU{$^{(9)}$}
\def\iCSU{$^{(10)}$}
\def\iFERR{$^{(11)}$}
\def\iFRAS{$^{(12)}$}
\def\iILLI{$^{(13)}$}
\def\iJHU{$^{(14)}$}
\def\iLBL{$^{(15)}$}
\def\iLTU{$^{(16)}$}
\def\iMASS{$^{(17)}$}
\def\iMISSI{$^{(18)}$}
\def\iMIT{$^{(19)}$}
\def\iMOSCOW{$^{(20)}$}
\def\iNAGO{$^{(21)}$}
\def\iOREG{$^{(22)}$}
\def\iOXF{$^{(23)}$}
\def\iPADO{$^{(24)}$}
\def\iPERU{$^{(25)}$}
\def\iPISA{$^{(26)}$}
\def\iRAL{$^{(27)}$}
\def\iRUTG{$^{(28)}$}
\def\iSLAC{$^{(29)}$}
\def\iSOGA{$^{(30)}$}
\def\iSOONG{$^{(31)}$}
\def\iTENN{$^{(32)}$}
\def\iTOHO{$^{(33)}$}
\def\iUCSB{$^{(34)}$}
\def\iUCSC{$^{(35)}$}
\def\iUVIC{$^{(36)}$}
\def\iVAND{$^{(37)}$}
\def\iWASH{$^{(38)}$}
\def\iWISC{$^{(39)}$}
\def\iYALE{$^{(40)}$}

  \baselineskip=.75\baselineskip  
\mbox{Kenji  Abe\unskip,\iNAGO}
\mbox{Koya Abe\unskip,\iTOHO}
\mbox{T. Abe\unskip,\iSLAC}
\mbox{I. Adam\unskip,\iSLAC}
\mbox{T.  Akagi\unskip,\iSLAC}
\mbox{H. Akimoto\unskip,\iSLAC}
\mbox{N.J. Allen\unskip,\iBRUN}
\mbox{W.W. Ash\unskip,\iSLAC}
\mbox{D. Aston\unskip,\iSLAC}
\mbox{K.G. Baird\unskip,\iMASS}
\mbox{C. Baltay\unskip,\iYALE}
\mbox{H.R. Band\unskip,\iWISC}
\mbox{M.B. Barakat\unskip,\iLTU}
\mbox{O. Bardon\unskip,\iMIT}
\mbox{T.L. Barklow\unskip,\iSLAC}
\mbox{G.L. Bashindzhagyan\unskip,\iMOSCOW}
\mbox{J.M. Bauer\unskip,\iMISSI}
\mbox{G. Bellodi\unskip,\iOXF}
\mbox{A.C. Benvenuti\unskip,\iBOLO}
\mbox{G.M. Bilei\unskip,\iPERU}
\mbox{D. Bisello\unskip,\iPADO}
\mbox{G. Blaylock\unskip,\iMASS}
\mbox{J.R. Bogart\unskip,\iSLAC}
\mbox{G.R. Bower\unskip,\iSLAC}
\mbox{J.E. Brau\unskip,\iOREG}
\mbox{M. Breidenbach\unskip,\iSLAC}
\mbox{W.M. Bugg\unskip,\iTENN}
\mbox{D. Burke\unskip,\iSLAC}
\mbox{T.H. Burnett\unskip,\iWASH}
\mbox{P.N. Burrows\unskip,\iOXF}
\mbox{R.M. Byrne\unskip,\iMIT}
\mbox{A. Calcaterra\unskip,\iFRAS}
\mbox{D. Calloway\unskip,\iSLAC}
\mbox{B. Camanzi\unskip,\iFERR}
\mbox{M. Carpinelli\unskip,\iPISA}
\mbox{R. Cassell\unskip,\iSLAC}
\mbox{R. Castaldi\unskip,\iPISA}
\mbox{A. Castro\unskip,\iPADO}
\mbox{M. Cavalli-Sforza\unskip,\iUCSC}
\mbox{A. Chou\unskip,\iSLAC}
\mbox{E. Church\unskip,\iWASH}
\mbox{H.O. Cohn\unskip,\iTENN}
\mbox{J.A. Coller\unskip,\iBU}
\mbox{M.R. Convery\unskip,\iSLAC}
\mbox{V. Cook\unskip,\iWASH}
\mbox{R.F. Cowan\unskip,\iMIT}
\mbox{D.G. Coyne\unskip,\iUCSC}
\mbox{G. Crawford\unskip,\iSLAC}
\mbox{C.J.S. Damerell\unskip,\iRAL}
\mbox{M.N. Danielson\unskip,\iCOLO}
\mbox{M. Daoudi\unskip,\iSLAC}
\mbox{N. de Groot\unskip,\iBRI}
\mbox{R. Dell'Orso\unskip,\iPERU}
\mbox{P.J. Dervan\unskip,\iBRUN}
\mbox{R. de Sangro\unskip,\iFRAS}
\mbox{M. Dima\unskip,\iCSU}
\mbox{D.N. Dong\unskip,\iMIT}
\mbox{M. Doser\unskip,\iSLAC}
\mbox{R. Dubois\unskip,\iSLAC}
\mbox{B.I. Eisenstein\unskip,\iILLI}
\mbox{I.Erofeeva\unskip,\iMOSCOW}
\mbox{V. Eschenburg\unskip,\iMISSI}
\mbox{E. Etzion\unskip,\iWISC}
\mbox{S. Fahey\unskip,\iCOLO}
\mbox{D. Falciai\unskip,\iFRAS}
\mbox{C. Fan\unskip,\iCOLO}
\mbox{J.P. Fernandez\unskip,\iUCSC}
\mbox{M.J. Fero\unskip,\iMIT}
\mbox{K. Flood\unskip,\iMASS}
\mbox{R. Frey\unskip,\iOREG}
\mbox{J. Gifford\unskip,\iUVIC}
\mbox{T. Gillman\unskip,\iRAL}
\mbox{G. Gladding\unskip,\iILLI}
\mbox{S. Gonzalez\unskip,\iMIT}
\mbox{E.R. Goodman\unskip,\iCOLO}
\mbox{E.L. Hart\unskip,\iTENN}
\mbox{J.L. Harton\unskip,\iCSU}
\mbox{K. Hasuko\unskip,\iTOHO}
\mbox{S.J. Hedges\unskip,\iBU}
\mbox{S.S. Hertzbach\unskip,\iMASS}
\mbox{M.D. Hildreth\unskip,\iSLAC}
\mbox{J. Huber\unskip,\iOREG}
\mbox{M.E. Huffer\unskip,\iSLAC}
\mbox{E.W. Hughes\unskip,\iSLAC}
\mbox{X. Huynh\unskip,\iSLAC}
\mbox{H. Hwang\unskip,\iOREG}
\mbox{M. Iwasaki\unskip,\iOREG}
\mbox{D.J. Jackson\unskip,\iRAL}
\mbox{P. Jacques\unskip,\iRUTG}
\mbox{J.A. Jaros\unskip,\iSLAC}
\mbox{Z.Y. Jiang\unskip,\iSLAC}
\mbox{A.S. Johnson\unskip,\iSLAC}
\mbox{J.R. Johnson\unskip,\iWISC}
\mbox{R.A. Johnson\unskip,\iCINC}
\mbox{T. Junk\unskip,\iSLAC}
\mbox{R. Kajikawa\unskip,\iNAGO}
\mbox{M. Kalelkar\unskip,\iRUTG}
\mbox{Y. Kamyshkov\unskip,\iTENN}
\mbox{H.J. Kang\unskip,\iRUTG}
\mbox{I. Karliner\unskip,\iILLI}
\mbox{H. Kawahara\unskip,\iSLAC}
\mbox{Y.D. Kim\unskip,\iSOGA}
\mbox{M.E. King\unskip,\iSLAC}
\mbox{R. King\unskip,\iSLAC}
\mbox{R.R. Kofler\unskip,\iMASS}
\mbox{N.M. Krishna\unskip,\iCOLO}
\mbox{R.S. Kroeger\unskip,\iMISSI}
\mbox{M. Langston\unskip,\iOREG}
\mbox{A. Lath\unskip,\iMIT}
\mbox{D.W.G. Leith\unskip,\iSLAC}
\mbox{V. Lia\unskip,\iMIT}
\mbox{C.Lin\unskip,\iMASS}
\mbox{M.X. Liu\unskip,\iYALE}
\mbox{X. Liu\unskip,\iUCSC}
\mbox{M. Loreti\unskip,\iPADO}
\mbox{A. Lu\unskip,\iUCSB}
\mbox{H.L. Lynch\unskip,\iSLAC}
\mbox{J. Ma\unskip,\iWASH}
\mbox{M. Mahjouri\unskip,\iMIT}
\mbox{G. Mancinelli\unskip,\iRUTG}
\mbox{S. Manly\unskip,\iYALE}
\mbox{G. Mantovani\unskip,\iPERU}
\mbox{T.W. Markiewicz\unskip,\iSLAC}
\mbox{T. Maruyama\unskip,\iSLAC}
\mbox{H. Masuda\unskip,\iSLAC}
\mbox{E. Mazzucato\unskip,\iFERR}
\mbox{A.K. McKemey\unskip,\iBRUN}
\mbox{B.T. Meadows\unskip,\iCINC}
\mbox{G. Menegatti\unskip,\iFERR}
\mbox{R. Messner\unskip,\iSLAC}
\mbox{P.M. Mockett\unskip,\iWASH}
\mbox{K.C. Moffeit\unskip,\iSLAC}
\mbox{T.B. Moore\unskip,\iYALE}
\mbox{M.Morii\unskip,\iSLAC}
\mbox{D. Muller\unskip,\iSLAC}
\mbox{V. Murzin\unskip,\iMOSCOW}
\mbox{T. Nagamine\unskip,\iTOHO}
\mbox{S. Narita\unskip,\iTOHO}
\mbox{U. Nauenberg\unskip,\iCOLO}
\mbox{H. Neal\unskip,\iSLAC}
\mbox{M. Nussbaum\unskip,\iCINC}
\mbox{N. Oishi\unskip,\iNAGO}
\mbox{D. Onoprienko\unskip,\iTENN}
\mbox{L.S. Osborne\unskip,\iMIT}
\mbox{R.S. Panvini\unskip,\iVAND}
\mbox{C.H. Park\unskip,\iSOONG}
\mbox{T.J. Pavel\unskip,\iSLAC}
\mbox{I. Peruzzi\unskip,\iFRAS}
\mbox{M. Piccolo\unskip,\iFRAS}
\mbox{L. Piemontese\unskip,\iFERR}
\mbox{K.T. Pitts\unskip,\iOREG}
\mbox{R.J. Plano\unskip,\iRUTG}
\mbox{R. Prepost\unskip,\iWISC}
\mbox{C.Y. Prescott\unskip,\iSLAC}
\mbox{G.D. Punkar\unskip,\iSLAC}
\mbox{J. Quigley\unskip,\iMIT}
\mbox{B.N. Ratcliff\unskip,\iSLAC}
\mbox{T.W. Reeves\unskip,\iVAND}
\mbox{J. Reidy\unskip,\iMISSI}
\mbox{P.L. Reinertsen\unskip,\iUCSC}
\mbox{P.E. Rensing\unskip,\iSLAC}
\mbox{L.S. Rochester\unskip,\iSLAC}
\mbox{P.C. Rowson\unskip,\iCOLU}
\mbox{J.J. Russell\unskip,\iSLAC}
\mbox{O.H. Saxton\unskip,\iSLAC}
\mbox{T. Schalk\unskip,\iUCSC}
\mbox{R.H. Schindler\unskip,\iSLAC}
\mbox{B.A. Schumm\unskip,\iUCSC}
\mbox{J. Schwiening\unskip,\iSLAC}
\mbox{S. Sen\unskip,\iYALE}
\mbox{V.V. Serbo\unskip,\iSLAC}
\mbox{M.H. Shaevitz\unskip,\iCOLU}
\mbox{J.T. Shank\unskip,\iBU}
\mbox{G. Shapiro\unskip,\iLBL}
\mbox{D.J. Sherden\unskip,\iSLAC}
\mbox{K.D. Shmakov\unskip,\iTENN}
\mbox{C. Simopoulos\unskip,\iSLAC}
\mbox{N.B. Sinev\unskip,\iOREG}
\mbox{S.R. Smith\unskip,\iSLAC}
\mbox{M.B. Smy\unskip,\iCSU}
\mbox{J.A. Snyder\unskip,\iYALE}
\mbox{H. Staengle\unskip,\iCSU}
\mbox{A. Stahl\unskip,\iSLAC}
\mbox{P. Stamer\unskip,\iRUTG}
\mbox{H. Steiner\unskip,\iLBL}
\mbox{R. Steiner\unskip,\iADEL}
\mbox{M.G. Strauss\unskip,\iMASS}
\mbox{D. Su\unskip,\iSLAC}
\mbox{F. Suekane\unskip,\iTOHO}
\mbox{A. Sugiyama\unskip,\iNAGO}
\mbox{S. Suzuki\unskip,\iNAGO}
\mbox{M. Swartz\unskip,\iJHU}
\mbox{A. Szumilo\unskip,\iWASH}
\mbox{T. Takahashi\unskip,\iSLAC}
\mbox{F.E. Taylor\unskip,\iMIT}
\mbox{J. Thom\unskip,\iSLAC}
\mbox{E. Torrence\unskip,\iMIT}
\mbox{N.K. Toumbas\unskip,\iSLAC}
\mbox{T. Usher\unskip,\iSLAC}
\mbox{C. Vannini\unskip,\iPISA}
\mbox{J. Va'vra\unskip,\iSLAC}
\mbox{E. Vella\unskip,\iSLAC}
\mbox{J.P. Venuti\unskip,\iVAND}
\mbox{R. Verdier\unskip,\iMIT}
\mbox{P.G. Verdini\unskip,\iPISA}
\mbox{D.L. Wagner\unskip,\iCOLO}
\mbox{S.R. Wagner\unskip,\iSLAC}
\mbox{A.P. Waite\unskip,\iSLAC}
\mbox{S. Walston\unskip,\iOREG}
\mbox{S.J. Watts\unskip,\iBRUN}
\mbox{A.W. Weidemann\unskip,\iTENN}
\mbox{E. R. Weiss\unskip,\iWASH}
\mbox{J.S. Whitaker\unskip,\iBU}
\mbox{S.L. White\unskip,\iTENN}
\mbox{F.J. Wickens\unskip,\iRAL}
\mbox{B. Williams\unskip,\iCOLO}
\mbox{D.C. Williams\unskip,\iMIT}
\mbox{S.H. Williams\unskip,\iSLAC}
\mbox{S. Willocq\unskip,\iMASS}
\mbox{R.J. Wilson\unskip,\iCSU}
\mbox{W.J. Wisniewski\unskip,\iSLAC}
\mbox{J. L. Wittlin\unskip,\iMASS}
\mbox{M. Woods\unskip,\iSLAC}
\mbox{G.B. Word\unskip,\iVAND}
\mbox{T.R. Wright\unskip,\iWISC}
\mbox{J. Wyss\unskip,\iPADO}
\mbox{R.K. Yamamoto\unskip,\iMIT}
\mbox{J.M. Yamartino\unskip,\iMIT}
\mbox{X. Yang\unskip,\iOREG}
\mbox{J. Yashima\unskip,\iTOHO}
\mbox{S.J. Yellin\unskip,\iUCSB}
\mbox{C.C. Young\unskip,\iSLAC}
\mbox{H. Yuta\unskip,\iAOMORI}
\mbox{G. Zapalac\unskip,\iWISC}
\mbox{R.W. Zdarko\unskip,\iSLAC}
\mbox{J. Zhou\unskip.\iOREG}

\it
  \vskip \baselineskip                   
  \centerline{(The SLD Collaboration)}   
  \vskip \baselineskip        
  \baselineskip=.75\baselineskip   
\iADEL
  Adelphi University, Garden City, New York 11530, \break
\iAOMORI
  Aomori University, Aomori , 030 Japan, \break
\iBOLO
  INFN Sezione di Bologna, I-40126, Bologna, Italy, \break
\iBRI
  University of Bristol, Bristol, U.K., \break
\iBRUN
  Brunel University, Uxbridge, Middlesex, UB8 3PH United Kingdom, \break
\iBU
  Boston University, Boston, Massachusetts 02215, \break
\iCINC
  University of Cincinnati, Cincinnati, Ohio 45221, \break
\iCOLO
  University of Colorado, Boulder, Colorado 80309, \break
\iCOLU
  Columbia University, New York, New York 10533, \break
\iCSU
  Colorado State University, Ft. Collins, Colorado 80523, \break
\iFERR
  INFN Sezione di Ferrara and Universita di Ferrara, I-44100 Ferrara, Italy, \break
\iFRAS
  INFN Lab. Nazionali di Frascati, I-00044 Frascati, Italy, \break
\iILLI
  University of Illinois, Urbana, Illinois 61801, \break
\iJHU
  Johns Hopkins University,  Baltimore, Maryland 21218-2686, \break
\iLBL
  Lawrence Berkeley Laboratory, University of California, Berkeley, California 94720, \break
\iLTU
  Louisiana Technical University, Ruston,Louisiana 71272, \break
\iMASS
  University of Massachusetts, Amherst, Massachusetts 01003, \break
\iMISSI
  University of Mississippi, University, Mississippi 38677, \break
\iMIT
  Massachusetts Institute of Technology, Cambridge, Massachusetts 02139, \break
\iMOSCOW
  Institute of Nuclear Physics, Moscow State University, 119899, Moscow Russia, \break
\iNAGO
  Nagoya University, Chikusa-ku, Nagoya, 464 Japan, \break
\iOREG
  University of Oregon, Eugene, Oregon 97403, \break
\iOXF
  Oxford University, Oxford, OX1 3RH, United Kingdom, \break
\iPADO
  INFN Sezione di Padova and Universita di Padova I-35100, Padova, Italy, \break
\iPERU
  INFN Sezione di Perugia and Universita di Perugia, I-06100 Perugia, Italy, \break
\iPISA
  INFN Sezione di Pisa and Universita di Pisa, I-56010 Pisa, Italy, \break
\iRAL
  Rutherford Appleton Laboratory, Chilton, Didcot, Oxon OX11 0QX United Kingdom, \break
\iRUTG
  Rutgers University, Piscataway, New Jersey 08855, \break
\iSLAC
  Stanford Linear Accelerator Center, Stanford University, Stanford, California 94309, \break
\iSOGA
  Sogang University, Seoul, Korea, \break
\iSOONG
  Soongsil University, Seoul, Korea 156-743, \break
\iTENN
  University of Tennessee, Knoxville, Tennessee 37996, \break
\iTOHO
  Tohoku University, Sendai 980, Japan, \break
\iUCSB
  University of California at Santa Barbara, Santa Barbara, California 93106, \break
\iUCSC
  University of California at Santa Cruz, Santa Cruz, California 95064, \break
\iUVIC
  University of Victoria, Victoria, British Columbia, Canada V8W 3P6, \break
\iVAND
  Vanderbilt University, Nashville,Tennessee 37235, \break
\iWASH
  University of Washington, Seattle, Washington 98105, \break
\iWISC
  University of Wisconsin, Madison,Wisconsin 53706, \break
\iYALE
  Yale University, New Haven, Connecticut 06511. \break

\rm
%

\end{center}

\begin{table}
\caption{Relative systematic errors on the 1997-98 measurement of $A_b$ ({\bf jet-charge}).}
\begin{center}
\begin{tabular}{lcc}
{\bf Error Source} & {\bf Variation} & $\delta A_b/A_b$ \\ \hline
 & & \\
\multicolumn{3}{l}{\it \underline{Self-Calibration}} \\ 
$\alpha_b$ statistics & $\pm$1$\sigma$ & 1.8\% \\
$\lambda_b$ Correlation & JETSET, HERWIG & 1.4\% \\
$P(Q_b)$ shape & Different shapes & 0.8\% \\
$\cos\theta$ shape of $\alpha_b$ & MC Shape $vs$ Flat & 0.4\% \\
Light Flavor  & 50\% of correction & 0.3\% \\
 & & \\
\multicolumn{3}{l}{\it \underline{Analysis}} \\
Tag Composition & Mostly $\epsilon_c$ & 0.3\% \\
Detector Modeling & Tracking eff.  & 2.4\% \\
 &and resolution& \\
 &corrections on/off& \\
Beam Polarization & $\pm$0.8\% & 0.8\% \\
QCD & $x_{QCD}$, $\alpha_s\pm 0.007$,& 0.6\% \\
    &$ 2^{nd}$ order terms &\\
Gluon Splitting & $\pm$100\% of JETSET& 0.2\% \\
$A_c$ & $0.67\pm 0.08$ & $<$0.1\% \\
$A_{bckg}$ & $0\pm 0.50$ & 0.2\% \\
 \hline
{\bf Total} & & {\bf 3.6\%}
\end{tabular}
\end{center}
\label{systab_jetq}
\end{table}
\vfill\eject

\begin{table}[h]
\begin{center}
\begin{tabular}{|l|c|c|} \hline
Source  & $\Delta A_b (1993-8)  $ & $\Delta A_c (1993-8) $  \\
\hline\hline 
Monte Carlo statistics                        & $\pm$.0022 & $\pm$.0104 \\
Tracking efficiency                        & $\pm$.0055 & $\pm$.0035 \\
Jet axis simulation                      & $\pm$.0013 & $\pm$.0016 \\
Background level                                & $\pm$.0082 & $\pm$.0306 \\ 
Background asymmetry                               & $\mp$.0027 & $\pm$.0142 \\
BR($Z^0 \rightarrow b\bar{b}$)                     & $\mp$.0004 & $\pm$.0006 \\
BR($Z^0 \rightarrow c\bar{c}$)                     & $\pm$.0008 & $\mp$.0094 \\
BR($b \rightarrow \mu^-$)                          & $\mp$.0035 & $\pm$.0034 \\
BR($b \rightarrow c \rightarrow \mu^+$)            & $\pm$.0039 & $\mp$.0038 \\
BR($b \rightarrow \bar{c} \rightarrow \mu^-$)      & $\pm$.0037 & $\pm$.0113 \\
BR($b \rightarrow \tau \rightarrow \mu^-$)         & $\pm$.0002 & $\pm$.0023 \\
BR($b \rightarrow J / \psi  \rightarrow \mu^{\pm}$)& $\pm$.0028 & $\pm$.0004 \\
BR($c \rightarrow \mu^+$)                          & $\pm$.0018 & $\mp$.0197 \\
$B^{\pm,0}$ leptonic spectrum- $D^{**}$ fraction  & $\pm$.0028 & $\pm$.0028 \\
$B_s$ leptonic spectrum- $D^{**}$ fraction      & $\pm$.0007 & $\pm$.0003 \\
$D$ leptonic spectrum                          & $\pm$.0037 & $\pm$.0006 \\
BR($B\rightarrow D{\bar D}$)                  & $\pm$.0027 &  $\pm$.0003 \\
L/D syst                                      &  $\pm$.0037  &  $\pm$.0032 \\
B tag calibration					      & $\pm$.0137 &  $\pm$.0487 \\
$B_{s}$ fraction in $b\bar{b}$ events           & $\pm$.0009 & $\mp$.0012 \\
$\Lambda_{b}$ fraction in $b\bar{b}$ events    & $\pm$.0018 & $\mp$.0007 \\
$b$ fragmentation                             & $\pm$.0013 & $\pm$.0014 \\
$c$ fragmentation                             & $\pm$.0025   & $\pm$.0118 \\
Aleph/Peterson B fragmentation                & $\pm$.0034  & $\pm~.0022$ \\
Polarization                                  & $\mp$.0087 & $\mp$.0051 \\
Gluon splitting                               & $\pm$.0022  &$\pm$.0022    \\
Other QCD                                 & $\pm$.0040 & $\pm$.0030 \\
$B$ mixing                                         & $\pm$.0105 & $<$.0001   \\
$B$ mixing (cascade) 				& $\pm$ .0003  & $\pm$.0041\\
\hline 
Total systematic error   & 0.0250     & 0.0670      \\
\hline
\end{tabular}
\caption{Systematic errors on $A_b$ and $A_c$ measurements from 1993-8 data ({\bf $\mu$ tag}).}
\label{systab_mu}
\end{center}
\end{table}

\vfill\eject

\begin{table}[h]
\begin{tabular}{|l|c|c|c|} \hline
Source  & Variations adopted & $\Delta A_b (1997) $ & $\Delta A_b (1998) $  \\
\hline\hline 
Monte Carlo statistics & weights $w_i$ variation& $\pm$.011 & $\pm$.007 \\
Tracking efficiency   & MC/data track multiplicity & $<$.001 & $<$.001 \\
Jet axis simulation  & smearing 10 mrad & $\pm$.017 & $\pm$.049 \\
Background level    & $\pm~15\%$   & $\pm$.005 & $\pm$.007 \\ 
Background asymmetry  & $\pm~40\%$   & $\mp$.003 & $\pm$.004 \\
BR($Z^0 \rightarrow b\bar{b}$) & $(21.73\pm~.09)\%$ & $<$.001 & $<$.001 \\
BR($Z^0 \rightarrow c\bar{c}$) & $(17.30\pm~.44)\%$  & $\pm$.002 & $\mp$.002 \\
BR($b \rightarrow e^-$) & $(11.06\pm~.19)\%$ & $\mp$.003 & $\pm$.005 \\
BR($b \rightarrow c \rightarrow e^+$) & $(8.02\pm~.32)\%$  & $\pm$.003 & $\mp$.008 \\
BR($b \rightarrow \bar{c} \rightarrow e^-$) & $(1.3\pm~0.5)\%$  & $\pm$.001 & $\pm$.003 \\
BR($b \rightarrow \tau \rightarrow e^-$) & $(0.472\pm~.075)\%$ & $<$.001 & $\pm$.001 \\
BR($b \rightarrow J / \psi  \rightarrow e^{\pm}$)& $(0.07\pm~.02)\%$ & $\pm$.002 & $\pm$.002 \\
BR($c \rightarrow e^+$) & $(9.8\pm~0.5)\%$  & $\pm$.004 & $\mp$.005 \\
$B^{\pm,0}$- $D^{**}$ fraction & $(23\pm~10)\%$ & $\pm$.003 & $\pm$.001 \\
$B_s$ -$D^{**}$ fraction & $(32\pm~10)\%$ & $\pm$.003 & $\pm$ .002 \\
$D$ lepton spectrum   & $ACCMM1^{+2}_{-3}$   & $\pm$.003 & $\pm$.006 \\
$B_{s}$ fraction in $b\bar{b}$ events & $.115\pm~.050$ & $\pm$.004 & $\mp$.010 \\
$\Lambda_{b}$ fraction  in $b\bar{b}$ events & $.072\pm~.030$   & $\pm$.002 & $\mp$.005 \\
$b$ fragmentation  & $\epsilon_b=.0045-.0075$  & $<$.001 & $\pm$.002 \\
$c$ fragmentation & $\epsilon_c=.045-.070$   & $<$.001 & $\pm$.001 \\
Aleph & reweighting & $\pm$.004  & $\pm$.004  \\
Polarization      &$P=0.733\pm ~0.0080$  & $\mp$.008 & $\mp$.009 \\
Second order QCD  & $\Delta_{QCD} uncertainties $    & $\pm$.004 & $\pm$.004 \\
$B$ mixing    & $\bar{\chi}=.1217\pm~.0046$     & $\pm$.010 & $\pm$.011   \\
$B$ mixing cascade    &     $\bar{\chi}=.1285\pm~.0071$   & $\pm$.001 & $\pm$.004 \\
\hline 
Total systematic error &   & 0.027      & 0.055      \\
\hline
\end{tabular}
\caption{Systematic errors on $A_b$ from 1997 and 1998 data ({\bf $e^{\pm}$ tag}).}
\label{systab_el}
\end{table}


\begin{thebibliography}{60}



\bibitem{ZVTOP} D. Jackson, Nucl. Instrum. Methods {\bf A388}, 247
(1997).


\bibitem{JETQ98} K. Abe $et\ al.$  (SLD Collaboration),
Phys. Rev. Lett. {\bf 81}, 942 (1998);


\bibitem{MORIOND99} N. deGroot, to appear in {\sl Proceedings of the 
XXXIV Rencontres de Moriond, Les Arcs, France, Mar 13-20 1999.}

\bibitem{LTAG98} K. Abe $et\ al.$  (SLD Collaboration), SLAC-PUB-7798,
Submitted to {\sl Phys. Rev. Lett.}

\bibitem{KTAG98} K. Abe $et\ al.$  (SLD Collaboration), SLAC-PUB-7959,
Submitted to {\sl Phys. Rev. Lett.}



\bibitem{KTAGNEW} K. Abe $et\ al.$ (SLD Collaboration), SLAC-PUB-8200,
contributed paper to the {\sl International Europhysics Conference on
High Energy Physics (EPS-HEP99), Tampere, Finland, July 15-21 1999.}

\bibitem{VTXCHG} K. Abe $et\ al.$ (SLD Collaboration), SLAC-PUB-8201,
contributed paper to the {\sl International Europhysics Conference on
High Energy Physics (EPS-HEP99), Tampere, Finland, July 15-21 1999.}

\bibitem{LEPEWWG} The LEP Collaborations, ALEPH, DELPHI, L3, OPAL,
the LEP Electroweak Working Group, and the SLD Heavy Flavor Group,
CERN-EP/99-15.

\bibitem{MNICH} The LEP EWWG fit value is $A_b = 0.912 \pm 0.025$ (Summer 99). 
See for example J. Mnich, Electroweak Summary talk at the {\sl International Europhysics Conference on
High Energy Physics (EPS-HEP99), Tampere, Finland, July 15-21 1999.}

\bibitem{MOENIG} Klaus M\oe nig, private communication.

\bibitem{FIELD} J. H. Field and D. Sciarrino, preprint UGVA-DPNC 1999/7-183;
{\tt hep-ex/9907018}.

\end{thebibliography}
\end{document}